\documentclass[letter]{aa}

\usepackage{graphicx}

\usepackage{txfonts}

\usepackage{xcolor}
\usepackage[utf8]{inputenc}

\definecolor{xlinkcolor}{cmyk}{1,0.6,0,0}

\usepackage[breaklinks=true, 
    bookmarks=false,         
     pdfnewwindow=true,      
     colorlinks=true,    
     linkcolor=xlinkcolor,     
     citecolor=xlinkcolor,     
     filecolor=xlinkcolor,  
     urlcolor=xlinkcolor,      
final=true
]{hyperref}

\defcitealias{saulder+2020}{S20}

\begin{document}

   \title{Living on the edge. A quantitative warning on boundary artifacts in the IllustrisTNG}

   \subtitle{}

\titlerunning{A quantitative warning on boundary artifacts in the IllustrisTNG}

   \author{Ana Mitra{\v s}inovi{\' c}
          \inst{}
         }

   \institute{Astronomical Observatory, Volgina 7, 11060 Belgrade, Serbia\\
              \email{amitrasinovic@aob.rs}
             }

   \date{Received September 15, 1996; accepted March 16, 1997}

  \abstract
  {Periodic boundary conditions (PBCs) are a practical necessity in cosmological simulations, but they can also introduce numerical artifacts. We quantified the prevalence of PBC-related artifacts in the IllustrisTNG using dark-matter-deprived galaxies (DMDGs) as tracers. We found that their occurrence scales inversely with simulation volume. We demonstrated that this excess population is spatially correlated with the box edges; the smallest TNG50 box is most affected by these problems. This is a unique and irreplaceable resource for studying galaxy structure. Manual inspection confirmed abrupt, unphysical mass-loss events coincident with boundary crossings. We also highlight the challenge of disentangling numerical artifacts from genuine tidally stripped galaxies in boundary-crossing clusters in TNG100. We conclude by recommending a set of mandatory sanity checks that include positional verification, mass history analysis, and even exclusion of the buffer zone near the edges. These strategies ensure the robustness of the scientific results derived from these invaluable simulations. }

   \maketitle

\section{Introduction}

Cosmological simulations of galaxy formation are a valuable contemporary tool for bridging the gap between theoretical knowledge and vast observational data \citep[for reviews of the ecosystem of cosmological simulations, see][]{Vogelsberger+2020NatRP...2...42V, Crain+vaddeVoort2023ARA&A..61..473C}. The real Universe is effectively infinite, whereas cosmological simulations model the large-scale structure of the Universe within a finite cubic volume. As a solution, simulations use periodic boundary conditions \citep[PBCs; e.g.,][]{Hockney+Eastwood1988, springel2005}. When a particle, galaxy (typically subhalo), or halo crosses one boundary of the simulation box, it reenters from the opposite side, as if the simulation volume tiles infinitely, ensuring that structures can evolve without artificial edges. This solution works very well in practice, but it is not perfect and has its pitfalls. 

The identification of halos and subhalos in cosmological simulations relies on specialized structure-finding algorithms, each with advantages and limitations. Highly efficient, in a computational sense, is the \texttt{SUBFIND} algorithm \citep{SUBFIND2001}, which is a configuration-space algorithm that operates by detecting gravitationally bound overdensities. This algorithm is relevant to our work because it is used to produce subhalo catalogs in IllustrisTNG. Studies comparing various structure-finding algorithms \citep[e.g.,][]{Knebe+2011MNRAS.415.2293K, Muldrew+2011MNRAS.410.2617M, ForouharMoreno+2025MNRAS.tmp.1440F} have shown, however, that \texttt{SUBFIND} becomes less reliable for subhalos in very dense environments relative to algorithms that are based on phase space \citep[e.g.,][]{Canas+2019MNRAS.482.2039C, Elahi+2019PASA...36...21E} or history space \citep[e.g.,][]{Han+2018MNRAS.474..604H, Mansfield+2024ApJ...970..178M}. These findings suggest that analogous errors can be expected for subhalos near the edges of the periodic simulation box, where parts of structures lie across the boundary wrap, and local density or binding energy calculations may be perturbed, resulting in multiple artificial subhalos of nonphysical origin (i.e., numerical artifacts). Although alternative algorithms have the potential to resolve some of the issues, they can be more computationally demanding and still require careful treatment of periodic wrapping. Thus, while the choice of (sub)halo finder can affect the prevalence of numerical artifacts, no method is entirely immune to issues arising at the edges of the simulation box.

The discovery of dark-matter-deprived galaxies (DMDGs) by \citet{vanDokkum2018,vanDokkum2019} has raised questions about the validity of the standard cosmological model. It soon became clear that these galaxies naturally occur as a result of tidal stripping \citep[e.g.,][]{ogiya2018, montes2020, jackson2021, maccio2021, Montero-Dorta+2024MNRAS.527.5868M, moreno2022, Ogiya2022, Contreras-Santos+2024A&A...690A.109C}. Although this formation mechanism is probably dominant, DMDGs can also form from tidally stripped material in the aftermath of a high-velocity head-on collision between two gas-rich galaxies \citep{Silk2019MNRAS.488L..24S, Shin+2020ApJ...899...25S, Lee+2021ApJ...917L..15L, vanDokkum+2022Natur, Ivleva+2024A&A...687A.105I, Lee+2024ApJ...966...72L}. Most studies that explored DMDGs in simulations have focused on the satellite galaxy population, reinforcing the idea that DMDGs are indeed satellites. Some observed \citep[e.g.,][]{Guo+2020NatAs...4..246G, Cameron+2023A&A} and even simulated \citep[e.g.,][]{Mitrasinovic+2023A&A...680L...1M} DMDGs can be found in isolation, however, far from more massive galaxies. In an effort to explore these isolated DMDGs, \citet[][henceforth \citetalias{saulder+2020}]{saulder+2020} serendipitously found an issue with the implementation of the PBCs that causes regular galaxies crossing the edge of the simulation box to become torn apart, unrelated to any (sub)halo-finding algorithm. Ruediger Pakmor and Volker Springel traced and identified the bug as a force-calculation error in \texttt{GADGET-2} \citep{springel2005} and its derivatives (namely, \texttt{GADGET-3}, \texttt{Arepo}, and \texttt{GIZMO}), and they promptly fixed the issue. The resolution of this issue will improve the robustness and precision of the upcoming simulations, but numerical artifacts related to this issue \citepalias{saulder+2020} or the particular subhalo-finding algorithm still exist in the publicly available IllustrisTNG data. It is important to note that while observational searches for DMDGs \citep[e.g.,][]{vanDokkum2018, vanDokkum2019} often focus on low-mass satellite galaxies, the artifacts identified by \citetalias{saulder+2020} were primarily massive galaxies, suggesting that numerical issues at the boundary can affect a wide range of galaxy masses. This motivates a broad search for these effects.

The widespread use of IllustrisTNG simulations, particularly the high-resolution TNG50 volume \citep{TNG50-1-2019, TNG50-2-2019}, for studying galaxy evolution necessitates a clear understanding of their numerical limitations. The TNG50 simulation offers an unparalleled combination of resolution and cosmological context, making it a cornerstone for studying galaxy formation and internal structure. Given current trends in numerical astrophysics and the fact that newer simulations (e.g., MillenniumTNG or TNG-Cluster) focus on enlarging the simulation box and not on improving particle resolution \citep[e.g.,][]{Bose+2023MNRAS.524.2579B, Hernandez-Aguayo+2023MNRAS.524.2556H, Pakmor+2023MNRAS.524.2539P, Nelson+2024-tngcluster}, it is highly unlikely that the unique TNG50 will become irrelevant anytime soon. Although previous work has identified the existence of artifacts caused by PBCs \citepalias{saulder+2020}, the prevalence and scaling of these artifacts with simulation volume have not been quantified. The small volume of TNG50, while enabling its high resolution (which offers an invaluable resource for in-depth studies of galaxy structure in a cosmological context), may also increase the rate of numerical artifacts. This is because the small volume inherently increases the frequency of periodic boundary crossings.

The motivation for this work therefore is to provide an urgent, quantitative assessment of this issue. We provide the first quantitative estimate of the prevalence of PBC-related artifacts, using DMDGs as a proxy, in the three main IllustrisTNG volumes (TNG50, TNG100, and TNG300). Our goal is to demonstrate how the prevalence of these artifacts scales with box size and to highlight the specific challenges and interpretational pitfalls that are primarily present in TNG50 but, to a certain extent, also in TNG100. By doing so, we aim to provide not only a necessary warning, but also a constructive guide with practical mitigation strategies to safeguard the robustness of current and future scientific analyses that rely on these invaluable simulations. This paper is organized as follows. In Sect.~\ref{sec:sims} we briefly introduce the simulations and describe our sample and proxies. In Sect.~\ref{sec:results} we present the results of our calculations, present an illustrative example of a numerical artifact, and analyze the optimal value of the buffer size by examining the fractions of DMDGs in these boundary regions. Finally, in Sect.~\ref{sec:conclusion}, we discuss our results and their broader implications and offer comments on mitigation strategies.

\section{Simulations}\label{sec:sims}

The IllustrisTNG\footnote{Publicly available at \url{https://www.tng-project.org/data/}.} cosmological simulation suite \citep{TNGmethods2017,TNGmethods2018,Nelson+2019ComAC}, performed using the code \texttt{Arepo} \citep{Springel2010AREPO}, includes three flagship runs\footnote{Excluding the recently released TNG-Cluster \citep{Nelson+2024-tngcluster}.} with different cubic volumes: TNG50, TNG100, and TNG300. These are the focus of this work. These different volumes (TNG50, TNG100, and TNG300) are named approximately after the side length of the simulation box (i.e., 50, 100, and 300 Mpc, respectively). In addition to differences in volume, the three boxes also differ in particle resolution; as mentioned above, the TNG50, being the smallest box, excels in high particle resolution. In contrast, the TNG300 has the lowest particle resolution of the three boxes, but compensates for this by offering a larger volume. The TNG100 \citep{Marinacci+2018, Naiman+2018, Nelson+2018, Pillepich+2018, Springel+2018} offers an optimal balance between volume and resolution, representing the main simulation of general use.

To investigate the prevalence of potential numerical artifacts, we first identified a sample of sufficiently resolved galaxies with stellar masses $M_\star > 10^{8.5}\; \mathrm{M_\odot}$ at the present redshift $z=0$ in the TNG50, TNG100, and TNG300. The imposed lower mass limit ensures that the subhalo finder can reliably detect these galaxies, regardless of resolution, thereby making the comparison between simulation boxes robust \citep[see][]{Onions+2012MNRAS.423.1200O}. We opted for a fixed lower stellar mass cut and did not use a fixed particle number because this would result in comparing three physically distinct galaxy populations with different mass ranges because the resolutions in different simulation boxes differ. Our approach ensures that we compared the same physical class of objects across all boxes. We used DMDGs as proxies for numerical artifacts and defined them as galaxies with a dark matter mass $M_\mathrm{DM}$ lower than twice the mass of the baryonic components (stars and gas), $M_\mathrm{B}$, that is, $M_\mathrm{DM}< 2\; M_\mathrm{B}$. This liberal criterion is designed to capture a broad range of potential outliers.

\section{Results}\label{sec:results}

Our analysis revealed a monotonic trend between the simulation box volume and the fraction of DMDGs. As shown in the "Full Box" row of Table~\ref{tab:buffer}, the total DMDG fraction scales inversely with box size. It decreases from 6.2\% in TNG50 to 2.3\% in TNG300. This clear inverse scaling with box size suggests that a significant population of these DMDGs might not be of purely astrophysical origin.

To test the hypothesis that these excess DMDGs are related to PBCs, we analyzed their spatial distribution. We first defined a buffer zone as the region within a fixed, absolute distance of $1\;\mathrm{cMpc}$ from any edge of the simulation box, resulting in a fixed absolute volume for the buffer zone. The results show a striking spatial correlation. In TNG50, a remarkable 15.5\% (51/329) of all identified DMDGs reside within this narrow buffer zone. This percentage is also fairly high in TNG100, at 12.7\% (183/1444), but it plummets in TNG300, where only 2.4\% (227/9535) of DMDGs are found. This dramatic decrease in the largest-volume simulation suggests that a substantial fraction of the DMDGs identified in smaller boxes are numerical artifacts resulting from PBC-related issues. The number of DMDGs in buffer zones of different sizes, normalized by the total number of galaxies in the same region, is given in Table~\ref{tab:buffer} and is discussed in Sect.~\ref{sec:boundary}.

The buffer zone of the fixed absolute volume represents a different fraction of the total volume in each simulation box, however. To ensure the robustness of our conclusion, we performed a second test using a relative distance criterion. That is, we defined a buffer zone as the region within 1\% of the box size from any edge of the box. This ensured that the buffer zone occupies a fixed fraction of the total simulation volume in all three boxes ($\sim6\%$). Even with this volume-normalized definition, the trend persisted. The fraction of DMDGs located in this relative buffer zone is highest in TNG50 at 9.4\% (31/329), decreases to 7.3\% (105/1444) in TNG100, and is lowest in TNG300 at 5.1\% (485/9535). Although the trend is less extreme because the absolute size of the buffer zone is much larger in TNG300 ($3\; \mathrm{cMpc}$) and smaller in TNG50 ($0.5\; \mathrm{cMpc}$), the monotonic decrease remains clear. We also note that the absolute size of the buffer zone is probably overestimated in TNG300 (as it is unlikely to be affected by PBC-related issues in its entirety), and it is similarly underestimated in TNG50. The fact that a higher fraction of DMDGs is located near the edges in smaller boxes, under both absolute and relative criteria, provides compelling evidence that these objects are, possibly in significant numbers, numerical artifacts.

\subsection{Visual example of a boundary-crossing artifact}

\begin{figure}[ht!]
\centering \includegraphics[width=\columnwidth, keepaspectratio]{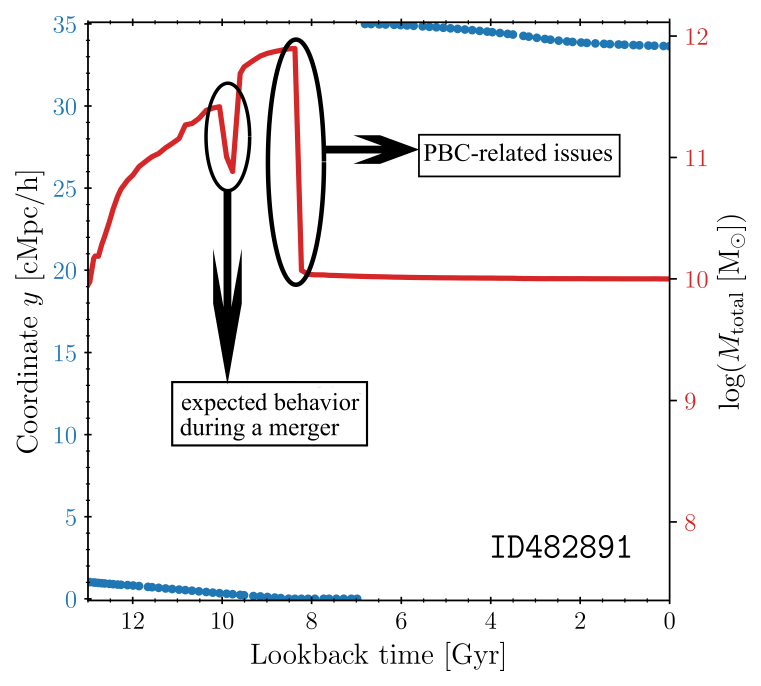}
\caption{Illustrative example of a numerical artifact in TNG50. The total mass evolution of the galaxy is shown by the red line (right y-axis), and its relevant coordinate $y$ is shown by the blue dots (left y-axis). A sudden, unphysical mass-loss event is clearly visible and annotated as a PBC-related issue. This event coincides with the moment in which the galaxy approached the boundary, slightly before the boundary-crossing event.}
\label{fig:example_single}
\end{figure}

In Fig.~\ref{fig:example_single} we show an example of a numerical artifact in TNG50, where the sudden drop in the total subhalo mass coincides with the boundary-crossing event. By visually and manually exploring all 329 DMDGs in TNG50 for similar features like in the provided example (i.e., searching for a sudden mass drop or almost instantaneous tidal stripping correlated with a boundary crossing or its proximity), we found that as many as 44 galaxies are numerical artifacts (nearly all that dwell near the edges because 51 DMDGs are found in the fixed absolute-volume buffer zone).

To determine a physically motivated buffer size, we first examined the positions at present-day redshift $z=0$ of the 44 visually confirmed artifacts in TNG50. We found that a significant fraction (more than half) indeed reside within $1-2$ cMpc/h of a boundary, confirming that proximity is a strong, though imperfect, indicator. The picture is complicated by the fact that these artifacts are not stationary, however. Based on their velocity and the time elapsed since the stripping event, the artifacts can migrate. We identified several clear artifacts at $z=0$ that are now about $3$ cMpc/h from the nearest boundary.

This artifact migration blurs the line for a single optimal buffer zone. A high-velocity galaxy that was unphysically stripped several billion years ago may have traveled deep into the "safe" region of the box. In contrast, a low-velocity galaxy that recently crossed a boundary might still be very close. This creates a trade-off: a small buffer (e.g., $1$ cMpc/h) reliably identifies recent or low-velocity artifacts, but misses those that migrated. Conversely, a very large buffer (e.g., 3 cMpc/h) might capture these older artifacts, but risks unnecessarily excluding a significant fraction of the box volume, and with it, many genuine physical galaxies. Therefore, we tested multiple absolute buffer sizes to show the scale of the contamination rather than relying on a single, arbitrary cut.

\subsection{A rigorous analysis of the boundary region}\label{sec:boundary}

To rigorously quantify the effect of this contamination, we tested a range of absolute buffer zones from $0.5$ cMpc/h to $3$ cMpc/h. For each buffer zone, we calculated the fraction of all mass-selected galaxies inside this specific zone that were classified as DMDGs. This provided a normalized, directly comparable metric of how strongly contaminated the different boundary regions are in each simulation. The results are presented in Table~\ref{tab:buffer}.

\begin{table*}[ht!]
\centering
\caption{Fractions of DMDGs and their number densities in three different simulation volumes (TNG50, TNG100, and TNG300) within buffer zones of various sizes.} \label{tab:buffer}

\begin{tabular}{|c|c|c|c|c|c|c|}
\hline
\begin{tabular}[c]{@{}c@{}}Buffer Size\\ (cMpc/h)\end{tabular} & \multicolumn{2}{c|}{TNG50} & \multicolumn{2}{c|}{TNG100} & \multicolumn{2}{c|}{TNG300} \\ \cline{2-7} 
 & \begin{tabular}[c]{@{}c@{}}$N_\mathrm{DMDG}/N_\mathrm{total}$ \\ (\%)\end{tabular} & \begin{tabular}[c]{@{}c@{}}Density\\ $(10\; \mathrm{cMpc}/h)^{-3}$\end{tabular} & \begin{tabular}[c]{@{}c@{}}$N_\mathrm{DMDG}/N_\mathrm{total}$ \\ (\%)\end{tabular} & \begin{tabular}[c]{@{}c@{}}Density\\ $(10\; \mathrm{cMpc}/h)^{-3}$\end{tabular} & \begin{tabular}[c]{@{}c@{}}$N_\mathrm{DMDG}/N_\mathrm{total}$ \\ (\%)\end{tabular} & \begin{tabular}[c]{@{}c@{}}Density\\ $(10\; \mathrm{cMpc}/h)^{-3}$\end{tabular} \\ \hline \hline

0.5 & 35/457 (7.7\%) & 9.80 & 55/1535 (3.6\%) & 3.30 & 94/5315 (1.8\%) & 0.75 \\ \hline
1.0 & 51/848 (6.0\%) & 7.35 & 183/3317 (5.5\%) & 5.57 & 227/10656 (2.1\%) & 0.91 \\ \hline
1.5 & 83/1191 (7.0\%) & 8.21 & 260/4850 (5.4\%) & 5.35 & 331/15896 (2.1\%) & 0.89 \\ \hline
2.0 & 86/1483 (5.8\%) & 6.57 & 314/6193 (5.1\%) & 4.91 & 471/21292 (2.2\%) & 0.95 \\ \hline
2.5 & 99/1795 (5.5\%) & 6.24 & 349/7279 (4.8\%) & 4.42 & 566/26607 (2.1\%) & 0.92 \\ \hline
3.0 & 111/2119 (5.2\%) & 6.00 & 364/8272 (4.4\%) & 3.90 & 676/31836 (2.1\%) & 0.92 \\ \hline \hline
Full Box & 329/5336 (6.2\%) & 7.67 & 1444/33423 (4.3\%) & 3.42 & 9535/412956 (2.3\%) & 1.11 \\ \hline
\end{tabular}
\tablefoot{Parameter $N_\mathrm{DMDG}$ is the number of DMDGs within the buffer zone, and $N_\mathrm{total}$ is the total number of galaxies in the same region. The lower stellar mass cut defined in Sect.~\ref{sec:sims} was applied to all galaxies.}
\end{table*}

In TNG300, the contamination rate is low and the background level is very stable at $\sim 2.1\%$, which is consistent with its overall average. We interpret this as the approximate physical rate of DMDG formation in the simulation. It is undeniable that artifacts might also form in TNG300, but their overall numbers are negligible. In stark contrast, the rate in TNG50 is elevated, peaks at $7.7\%$ in the $0.5$ cMpc/h buffer, and remains significantly higher than the TNG300 baseline in all tested zones. This demonstrates that a galaxy within the boundary region of TNG50 is three to four times more likely to be a DMDG than a galaxy in TNG300. The TNG100 data also show a consistently elevated rate (peaks at $5.5\%$), which supports our discussion (see Sect.~\ref{sec:conclusion}) that this is a mixed population of artifacts and genuine cluster-stripped DMDGs. Although this calculation did not provide a clear optimal value for the size of the buffer, the results presented in Table~\ref{tab:buffer} as well as the previously discussed visual inspection imply that a buffer of $\sim 1.5-2$ cMpc/h is a well-motivated choice to identify the most heavily contaminated region.

\section{Discussion and conclusion}\label{sec:conclusion}

Our results provide quantitative evidence for a significant box-size-dependent population of numerical artifacts in IllustrisTNG that mimic DMDGs. The prevalence of these artifacts is most pronounced in the TNG50 simulation, where almost all DMDGs that reside near the edges are numerical artifacts. Thus, the primary implication of our results is that TNG50, while a unique resource for studying galaxy structure, is the most susceptible to this systematic effect, which can produce artifacts that convincingly mimic real astrophysical phenomena, such as extreme tidal stripping.

An interesting feature of our results is the relatively modest decrease in the fraction of edge-associated DMDGs between TNG50 and TNG100, particularly compared to the dramatic drop observed in TNG300. We posit that this is due to the specific large-scale structure configuration in TNG100 at $z=0$, where several massive galaxy clusters are bisected by periodic boundaries. This particular configuration presents a unique and significant challenge for interpretation. Galaxy clusters are precisely the environments in which a higher incidence of physically formed DMDGs is expected because of the frequent gravitational interactions and intense tidal stripping. Consequently, the population of DMDGs near the edges of TNG100 is likely a complex mixture of two spatially coincident populations: numerical artifacts generated by PBC-related issues, and physically genuine tidally stripped galaxies that happen to reside in a boundary-crossing cluster. This conflation of numerical and physical effects means that extreme caution is advised for studying environmental processes in these specific clusters in TNG100. Hence, while PBC-related artifacts contaminate galaxy samples, they do not preclude the existence of real DMDGs in these same regions, making careful vetting indispensable. 

Furthermore, we have used DMDGs as proxies of the most obvious primary numerical artifacts (those with unphysically high mass loss). This lost mass does not simply vanish, however. It is often incorrectly assigned to nearby subhalos by the structure-finding algorithm. This creates a population of secondary artifacts: otherwise possibly normal-looking galaxies that have been artificially enhanced in mass. This implies that the total number of subhalos contaminated by PBC effects is likely at least double the number of unphysical DMDGs we quantified, making this a more prevalent issue than the analysis of a single tracer population would suggest.

Based on these findings, we propose a set of mitigation strategies as best practices for studies using IllustrisTNG, especially when using TNG50 or exploring boundary-crossing clusters in TNG100. We strongly recommend that researchers perform mandatory sanity checks on any objects of interest. These checks should include a positional verification to determine the proximity of a candidate galaxy to the box edges and a historical analysis to examine its mass and orbital history for unphysical events or discontinuities correlated with boundary crossings. For statistical studies, researchers should consider excluding all objects within a defined buffer zone (e.g., $1-2\; \mathrm{cMpc}$ from an edge) and test whether their conclusions are robust to this exclusion.

In conclusion, a critical understanding of the limitations of a simulation is crucial to maximize its scientific value. By quantifying the prevalence of PBC-related artifacts and providing clear mitigation strategies, we hope to enable more robust scientific analyses and strengthen the conclusions drawn from the invaluable IllustrisTNG simulation suite.

\begin{acknowledgements}
We are grateful to the IllustrisTNG team for making their simulations publicly available. We also thank the reviewer for their thoughtful and constructive feedback, which helped significantly improve the clarity, structure, and impact of a previous version of this manuscript. This research was supported by the Ministry of Science, Technological Development and Innovation of the Republic of Serbia (MSTDIRS) through contract no. 451-03-136/2025-03/200002, made with the Astronomical Observatory (Belgrade, Serbia).
\end{acknowledgements}

\bibliographystyle{aa} 
\bibliography{aa57557-25}

\end{document}